\documentclass{article}
\usepackage{spconf,amsmath,graphicx}
\usepackage{booktabs}
\usepackage{multirow}
\usepackage{enumitem}
\setlist{nosep, leftmargin=14pt}

\usepackage{mwe} 

\title{Iterative Data Refinement for Self-Supervised MR Image Reconstruction}
\name{Xue Liu$^{1,2}$, Juan Zou$^{1}$, Xiawu Zheng$^{3}$, Cheng Li$^{1}$, Hairong Zheng$^{1, 2}$, Shanshan Wang$^{1, 2}$}

\address{
$^{1}$Paul C. Lauterbur Research Center for Biomedical Imaging, Shenzhen Institutes of \\
Advanced Technology, Chinese Academy of Sciences, Shenzhen, Guangdong, China\\
$^{2}$University of Chinese Academy of Sciences, Beijing, China\\
$^{3}$Peng Cheng Laboratory, Shenzhen, Guangdong, China
}
\begin{document}
%
\maketitle
\begin{abstract}
Magnetic Resonance Imaging (MRI) has become an important technique in the clinic for the visualization, detection, and diagnosis of various diseases. However, one bottleneck limitation of MRI is the relatively slow data acquisition process. Fast MRI based on k-space undersampling and high-quality image reconstruction has been widely utilized, and many deep learning-based methods have been developed in recent years. Although promising results have been achieved, most existing methods require fully-sampled reference data for training the deep learning models. Unfortunately, fully-sampled MRI data are difficult if not impossible to obtain in real-world applications. To address this issue, we propose a data refinement framework for self-supervised MR image reconstruction. Specifically, we first analyze the reason of the performance gap between self-supervised and supervised methods and identify that the bias in the training datasets between the two is one major factor. Then, we design an effective self-supervised training data refinement method to reduce this data bias. With the data refinement, an enhanced self-supervised MR image reconstruction framework is developed to prompt accurate MR imaging. We evaluate our method on an in-vivo MRI dataset. Experimental results show that without utilizing any fully sampled MRI data, our self-supervised framework possesses strong capabilities in capturing image details and structures at high acceleration factors.
\end{abstract}
\section{Introduction}
\label{sec:intro}

Magnetic Resonance Imaging (MRI) is a critical imaging technique in the clinic. It is non-invasive and radiation-free. In the meantime, it can provide excellent soft-tissue contrast. Nevertheless, the speed of MRI data acquisition is slow, which not only increases patient discomfort but also introduces motion artifacts. One prevalent group of methods for MRI acceleration is k-space undersampling followed by high-quality MR image reconstruction. Here, the reconstruction is a highly ill-posed inverse problem. Traditional solutions are mainly based on the Compressive Sensing (CS) theory \cite{lustig2008compressed,ye2019compressed}. Methods such as Discrete Wavelet Transform (DWT) \cite{qu2012undersampled} and Total Variation (TV) \cite{block2007undersampled} have achieved promising reconstruction performance. However, these CS-based methods enhance the image quality in a step-by-step manner, which takes a long reconstruction time. Moreover, there are many hyper-parameters that require manual adjustment.

Recently, deep learning-based MR image reconstruction methods have achieved inspiring results\cite{li2022artificial,wang2021deep,wang2021review}. These methods can be roughly categorized into two types, data-driven and model-based approaches. Data-driven approaches apply neural networks to directly learn the mapping between undersampled and fully sampled data. Representative methods include RAKI \cite{akccakaya2019scan}, KIKI-Net \cite{eo2018kiki}, etc \cite{schlemper2017deep}. Model-based approaches unroll CS algorithms to construct the neural network structures. Some typical examples are ISTA-Net \cite{zhang2018ista}, ADMM-Net \cite{sun2016deep} and MoDL \cite{aggarwal2018modl}, etc \cite{hammernik2018learning}. Despite the performance achieved, these deep learning-based MR image reconstruction methods are supervised learning methods. They require fully sampled data to supervise the training process. In many real-world application scenarios, acquiring fully sampled data is difficult due to physiological and physical limitations. To solve this problem, unsupervised approaches have been proposed. The self-supervised learning method (SSDU) proposed by Yaman et al. \cite{yaman2020self} divides the acquired undersampled k-space data into two disjoint subsets. One subset is treated as the network input while the other is used to calculate the loss function. Then, the neural network can be trained without utilizing any fully sampled data. However, there exists a data bias in the training data of fully supervised models and self-supervised models, which leads to SSDU’s deteriorated reconstruction performance when compared to the fully supervised counterparts. To this end, we propose a  training data refinement method based on SSDU. Specifically, our method iteratively improve the quality of the training data by feeding the training target of the current iteration into the network and generate the refined training target for the next iteration. In this way, the bias in the training data can be gradually eliminated, and better reconstruction results are expected. The main contributions of this paper are as follows:
\begin{itemize}
\item A method is proposed to iteratively improve the quality of the training data and thus, reduce the data bias caused by the self-supervised learning data settings.
\item With our proposed training data refinement method, a more effective self-supervised learning framework is constructed to capture inherent features and details for undersampled MR image reconstruction.
\item The reconstruction performance of our framework is compared with state-of-the-art self-supervised MR image reconstruction approach utilizing an in vivo brain MRI dataset. Our framework achieves encouraging performances. 
\end{itemize}

\section{Method}
\label{sec:method}

\begin{figure*}[htbp]
\centering
\centerline{\includegraphics[width=14cm]{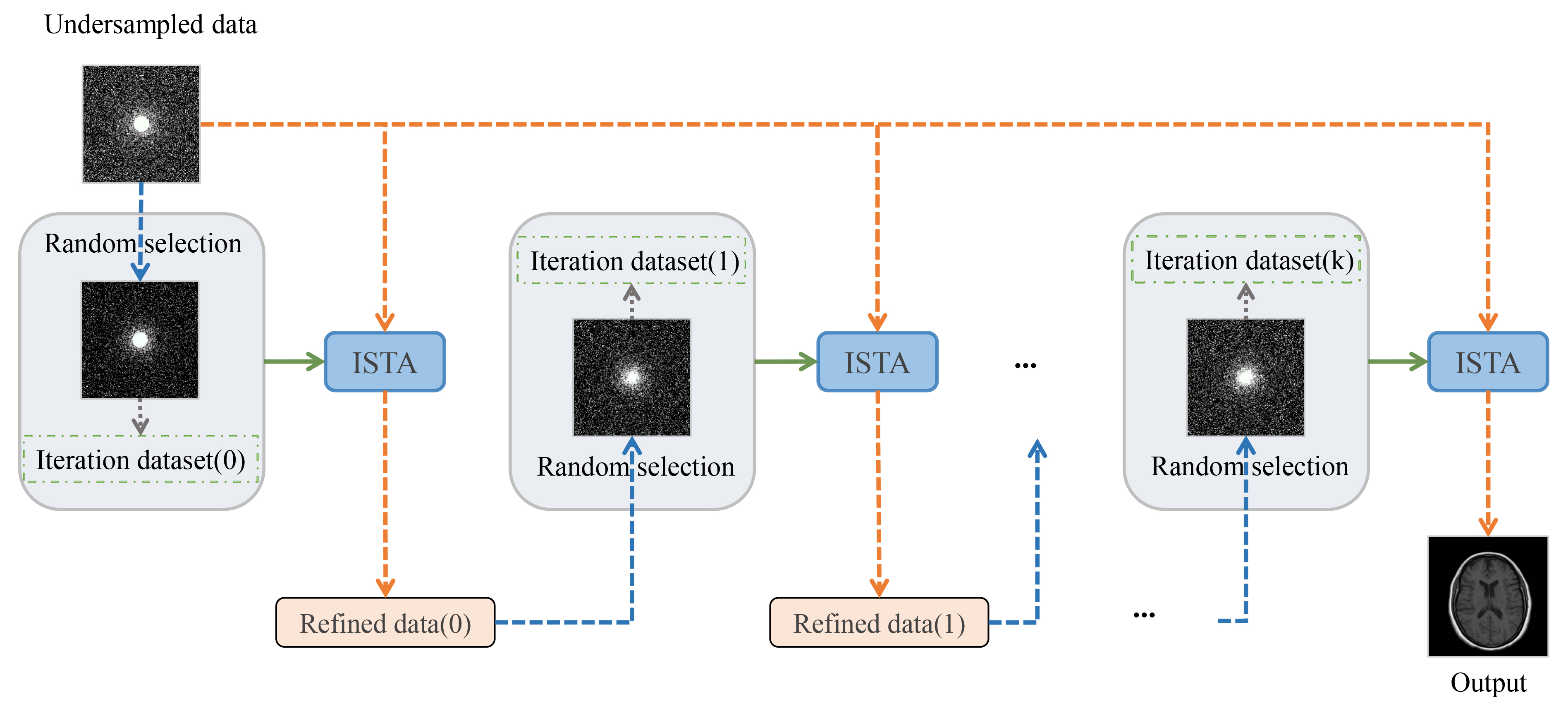}}
\caption{The pipeline of our proposed self-supervised MR image reconstruction framework embedding a novel training data refinement method.}
\label{fig:frame}
\end{figure*}

\subsection{Training Data Bias between Supervised and Self-Supervised Learning Framework}
\label{ssec:subme1}

The CS-MRI reconstruction problem can be written as:
$$
\arg \min _x \frac{1}{2}\left\|y_{\Omega}-E_{\Omega} x\right\|_2^2+\lambda R(x)
$$
where $x$ is the image to be reconstructed, $y_{\Omega}$ is the acquired k-space measurement with the undersampling mask $\Omega$, $E_{\Omega}$ denotes the encoding operator including Fourier transform F and sampling matrix $\Omega$, $R(x)$ donates the utilized regularization, and $\lambda$ is a constant parameter to control the contribution of $R(x)$.
When we train an end-to-end neural network in a supervised learning fashion, the fully sampled data is used to calculate the loss function to supervise the learning process. The loss function can be written as:
$$
\frac{1}{N} \sum_i^N \mathcal{L}\left(x^i, f\left(y_{\Omega}^i\right)\right)
$$
where $N$ is the number of training data, $x^i$ is the fully sampled reference image for subject $i$, $y_{\Omega}^i$ is the acquired undersampled k-space measurement with the undersampling mask $\Omega$ for subject i, $f(.)$ is the reconstruction network, $f(y_{\Omega}^i)$ denotes the reconstruction image, $L(.)$ denotes the loss calculated between the fully sampled image and the network reconstruction image. This loss formula is written in the image domain, it can also be written in the k-space:
$$
\frac{1}{N} \sum_i^N \mathcal{L}\left(x^i, E^if\left(y_{\Omega}^i\right)\right)
$$
where $y^i$ is the fully sampled k-space reference data, and $E^i$ is the encoding operator that transforms the reconstruction image $f(y_{\Omega}^i)$  to the corresponding k-space data. The training dataset for supervised learning can be written as:
$$
\text { Supervised training dataset: }\left\{y_{\Omega}^i, y^i\right\}_{i=1}^N
$$
When fully sampled data are not available, existing self-supervised methods generally divide the undersampled data into two disjoint sets, with one of them treated as the input to the network and the other or the entire undersampled data utilized to calculate the loss. The loss function for existing self-supervised learning methods can be written as:
$$
\frac{1}{N} \sum_i^N \mathcal{L}\left(y_{\Omega}^i, E_{\Omega}^i f\left(y_{\Lambda \times \Omega}^i\right)\right)
$$
where $\Lambda$ refers to the random mask for subset data selection. The training dataset for self-supervised learning can be written as:
$$
\text{ Self-supervised training dataset: }
\left\{\tilde{y}_{\Lambda}^i, \tilde{y}^i\right\}_{i=1}^N, \text{where } \tilde{y}^i=y_{\Omega}^i
$$
According to the descriptions, it can be noticed that there is a bias in the two datasets. Namely, the supervised dataset covers more comprehensive frequencies and directional information, while the self-supervised dataset consists of undersamped information. This data bias introduces big challenges for the self-supervised learning framework, which should be designed to recover and capture more inherent features to remove this data bias issue. Our motivation for this work is to build enhanced self-supervised learning framework for MR image reconstruction by reducing the data bias existed in the two training framework.

\subsection{Self-Supervised MR Image Reconstruction Framework with Iterative Data Refinement}
\label{ssec:subme2}

In our proposed framework, the network is trained in a stage-by-stage manner with the training data refined in each stage.

In the first step, we train a model with the initial self-supervised dataset $\left\{\tilde{y}_{\Lambda}^i, \tilde{y}^i\right\}$. Here, $\tilde{y}^i$ is the acquired undersampled data, and $\tilde{y}_{\Lambda}^i$  is selected from $\tilde{y}^i$ using a random mask $\Lambda$. The mask $\Lambda$ should be designed with the following criteria: 1) $\Lambda$ should select the data in a random manner which means they should be different with others. 2) The selected network input should include the majority of the low-frequency data points as well as a portion of the high-frequency data points in $\tilde{y}^i$. 3) We keep the number of selected data points to be roughly half of the number of the acquired undersampled data points. The entire acquired undersampled data points $\tilde{y}^i$ are used to calculate the reconstruction loss. With these settings, the training process is then divided into different iterative stages, in each stage the network is trained for several epochs. The best model $f_{0}$ in this first training stage is saved and adopted to refine the training data. Specifically, $\tilde{y}^i$ is inputted into $f_{0}$, and the output $f_{0}\left(\tilde{y}^i\right)$ is utilized as the refined data for the network training in the subsequent stage, i.e. the refined training dataset is $\left\{{f_{0}\left(\tilde{y}^i\right)}_{\Lambda},f_{0}\left(\tilde{y}^i\right)\right\}$. We keep the scanned data points unchanged during this data refinement process. Since $f_{0}\left(\tilde{y}^i\right)$ is reconstructed from $\tilde{y}^i$, it should be closer to the fully sampled data ${y}^i$. In this way, the bias is reduced.

In the second stage, $\left\{{f_{0}\left(\tilde{y}^i\right)}_{\Lambda},f_{0}\left(\tilde{y}^i\right)\right\}$ is used to train the network in a similar way to that of the first stage. After this stage, an updated model $f_{1}$ is saved to generate the refined training data $\left\{{f_{1}\left(\tilde{y}^i\right)}_{\Lambda},f_{1}\left(\tilde{y}^i\right)\right\}$. Then we can move on to the next stage. The whole network training and training data refinement process is depicted in Fig. 1. By iteratively refining the training data in each stage, a much better self-supervised training dataset can be obtained. Assume the data is refined for k stages, the final training dataset becomes $\left\{{f_{k-1}\left(\tilde{y}^i\right)}_{\Lambda},f_{k-1}\left(\tilde{y}^i\right)\right\}$. Ideally, with enough refinement stages, $f_{k-1}\left(\tilde{y}^i\right)$ should be infinitely close to the fully sampled data ${y}^i$, and the bias between the self-supervised training dataset and the supervised training dataset can be eliminated. Therefore, an enhanced self-supervised MR image reconstruction model $f_{k}$, which is gradually finetuned using the iteratively refined self-supervised training dataset from $\left\{\tilde{y}_{\Lambda}^i, \tilde{y}^i\right\}$to$\left\{{f_{k-1}\left(\tilde{y}^i\right)}_{\Lambda},f_{k-1}\left(\tilde{y}^i\right)\right\}$, can be obtained.

Overall, the network training process solves the following optimization problem:
$$
\arg \min _{x_j} \frac{1}{2}\left\|y_j-E_j x_j\right\|_2^2+\lambda R\left(x_j\right)
$$

$j$ is the stage number. $y_{j}$ is the selected undersampled k-space data input in each stage. $E_{j}$ is the corresponding encoding matrices. $x_{j}$ is the estimation of the ground-truth image $x$. $R\left(.\right)$ is the regularization. $\lambda$ is the weight constant. In the test phase, we employ the final best model $f_{k}$ to do the inference.

\subsection{Loss Function}
\label{ssec:subme3}

We chose ISTA \cite{zhang2018ista} as the base network structure in our proposed framework. In each iteration, the loss function can be written as:
$$
\mathcal{L}(\Theta)=\frac{1}{N} \sum_i^N \mathcal{L}\left(y_{\Omega}^i, E_{\Omega}^i f\left(y_{\Lambda \times \widetilde{\Omega}}^i, E_{\Lambda \times \widetilde{\Omega}}^i, \theta\right)+\gamma \mathcal{L}_{\text {cons }}\right)
$$
where $N$ is the number of training subjects, and $i$ denotes the $i^{th}$ training subject. $y_{\Omega}^i$ is the acquired k-space measurement with the undersampling mask $\Omega$ for subject $i$. $f(.)$ is the reconstruction network. $f\left(y_{\Lambda \times \widetilde{\Omega}}^i, E_{\Lambda \times \widetilde{\Omega}}^i\right)$ denotes the reconstruction image. $y_{\Lambda \times \widetilde{\Omega}}^i$ is the selected undersampled k-space data during each iteration, and $E_{\Lambda \times \widetilde{\Omega}}^i$ is the corresponding encoding matrices. In the first training stage, $\widetilde{\Omega}$ is the same as $\Omega$. Then, $\widetilde{\Omega}$ changes as the training data is refined. However, we do not know the exact undersampling mask during the iterative process. Thus, we use several existing undersampling masks to simulate the undersampling process. $\mathcal{L}_{\text {cons }}$ is the special constraint loss introduced in \cite{yaman2020self}. $L(.)$ is set as MSE loss. $\gamma$ is the regularization parameter, which is set to 0.01.

\section{EXPERIMENTS AND RESULT}
\label{sec:experiment}

\subsection{Dataset}
\label{ssec:dataset}

The brain data from the NYK fastMRI \cite{zbontar2018fastmri} database was used to test the reconstruction performance of different methods in this study. Only two-dimensional slices of T1 MR with an image size of 256×256 were used for the experiments in this paper. From this, 1200, 400, and 400 slices were chosen independently for the training, test, and validation sets, respectively.

\subsection{Implementation Details}
\label{ssec:implementation}

The proposed method is implemented in PyTorch. We use Xavier to initialize the network parameters with a gain of 1.0. To train the networks, we use Adam optimization with an initial learning rate of 0.001 and a batch size of 4. The learning rate is automatically reduced by a constant factor when the performance metric plateaus on the validation set. In our proposed framework, we train the network for 20 epochs in each stage in the default setting. There are 15 iteration stages in the training. But if the performance metric doesn’t change for more than 10 epochs, the current training stage will end immediately. All experiments are conducted on an Ubuntu 18.04 LTS (64-bit) operating system utilizing one NVIDIA TITAN Xp GPUs (each with a memory of 12 GB).

\subsection{Comparison Methods}
\label{ssec:compare}

\begin{table}[htbp]
\begin{tabular}{@{}ccccc@{}}
\toprule
\multirow{2}{*}{Methods} & \multicolumn{2}{c}{PSNR} & \multicolumn{2}{c}{SSIM} \\ \cmidrule(l){2-5} 
                         & 4$\times$   & 8$\times$  & 4$\times$   & 8$\times$  \\ \cmidrule(l){1-5}
U-Net-256                & 33.9010     & 32.4281    & 0.9164      & 0.9005     \\
SSDU                     & 45.8989     & 40.7511    & 0.9829      & 0.9659     \\
Ours                     & 46.9585     & 42.7554    & 0.9855      & 0.9752     \\
Supervised-ISTA          & 47.9833     & 43.0119    & 0.9891      & 0.9778     \\ \bottomrule
\end{tabular}
\caption{Quantitative results of different methods at two acceleration rates (4× and 8×).}
\label{table:res}
\end{table}

The MR image reconstruction performance of our proposed framework is compared to the following methods, a U-Net model trained in a supervised manner (U-Net-256), a self-supervised learning version of ISTA (SSDU), and a supervised learning version of ISTA (Supervised-ISTA). Peak signal-to-noise ratio (PSNR) and structural similarity index measurement (SSIM) are calculated to evaluate the performance. Table 1 lists the quantitative results. It can be observed that at both acceleration factors (4× and 8×) and for both metrics, our proposed method can always achieve better reconstruction results than the other two methods (U-Net-256 and SSDU). Additionally, the reconstruction performance of our self-supervised framework is comparable to that of the supervised method (Supervised-ISTA), which confirms the effectiveness of our proposed training data refinement method.

 Fig. 2 plots the reconstruction images of different methods as well as the corresponding error maps. Similar conclusions can be made to those with the quantitative results. Our framework generates more fine detailed structures than the other comparison methods (U-Net-256 and SSDU). The reconstruction errors of our framework are also smaller than two comparison methods. On the other hand, both the reconstruction images and the error maps of our framework look similar to those of the supervised method.

\begin{figure}[htbp]
\centering
\centerline{\includegraphics[width=8.5cm]{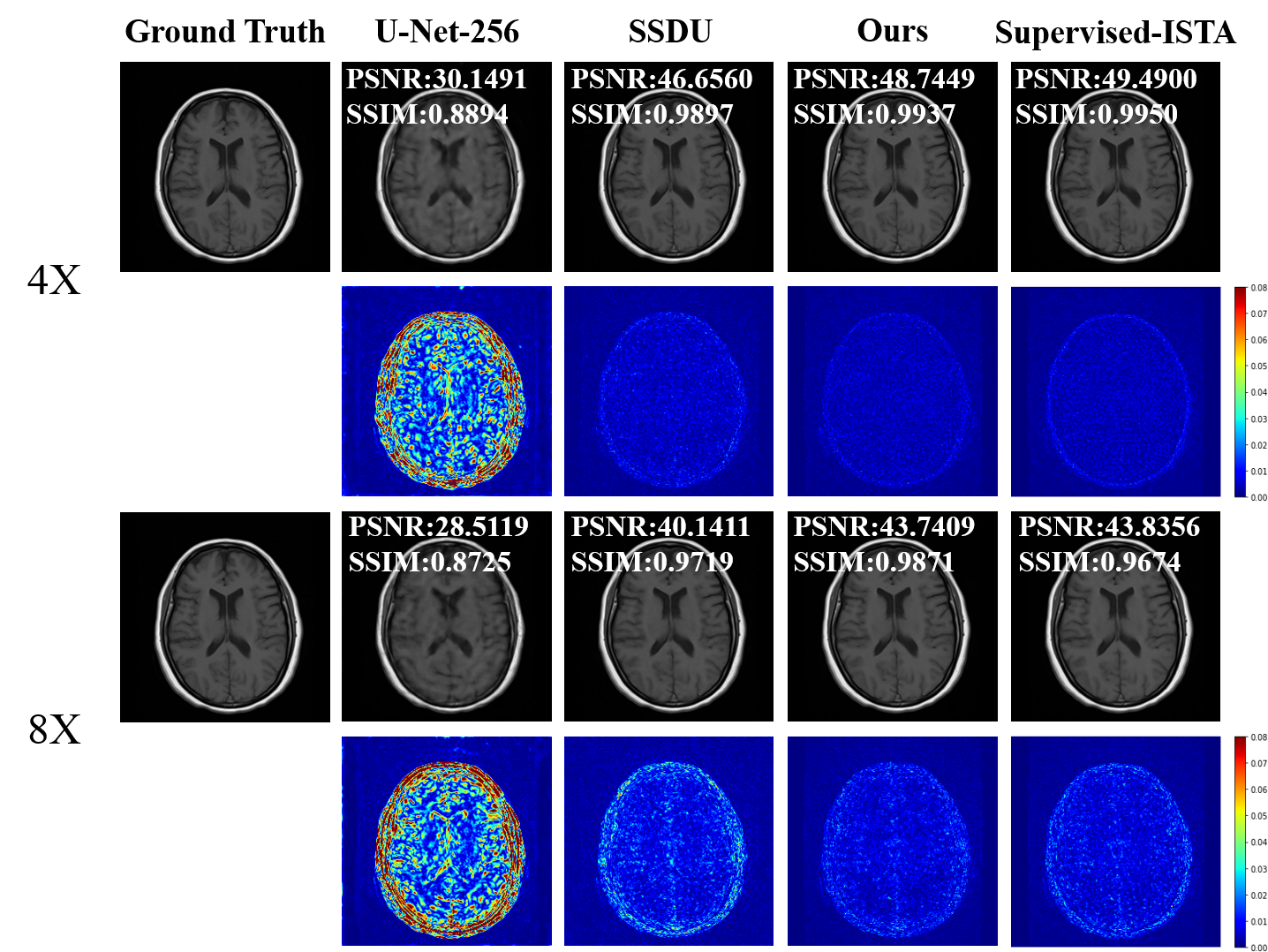}}
\caption{Example reconstruction results and their corresponding error maps of different methods at two acceleration rates (4× and 8×).}
\label{fig:res}
\end{figure}

\section{CONCLUSION}
\label{sec:conclusion}

In this paper, we propose a data refinement framework for self-supervised MR image reconstruction. With the gradually refined self-supervised training dataset, our framework can achieve enhanced MR image reconstruction without requiring fully sampled reference data during model training. Experiments on an in vivo dataset demonstrate that our framework achieves better reconstruction results when compared to existing self-supervised method. Thanks to the effectively reduced data bias, our framework can perform comparably to the supervised counterpart.


\section{Acknowledgments}
\label{sec:acknowledgments}

This research was partly supported by Scientific and Technical Innovation 2030-”New Generation Artificial Intelligence” Project (2020AAA0104100, 2020AAA0104105), the National Natural Science Foundation of China (61871371), Guangdong Provincial Key Laboratory of Artificial Intelligence in Medical Image Analysis and Application (Grant No. 2022B1212010011), the Basic Research Program of Shenzhen (JCYJ20180507182400762), Shenzhen Science and Technology Program (Grant No. RCYX20210706092104034), Youth Innovation Promotion Association Program of Chinese Academy of Sciences (2019351), and the Key Technology and Equipment R \& D Program of Major Science and Technology Infrastructure of Shenzhen: 202100102 and 202100104. 

\bibliographystyle{IEEEbib}
\bibliography{isbi}

\begin{thebibliography}{10}
\providecommand{\url}[1]{#1}
\csname url@samestyle\endcsname
\providecommand{\newblock}{\relax}
\providecommand{\bibinfo}[2]{#2}
\providecommand{\BIBentrySTDinterwordspacing}{\spaceskip=0pt\relax}
\providecommand{\BIBentryALTinterwordstretchfactor}{4}
\providecommand{\BIBentryALTinterwordspacing}{\spaceskip=\fontdimen2\font plus
\BIBentryALTinterwordstretchfactor\fontdimen3\font minus
  \fontdimen4\font\relax}
\providecommand{\BIBforeignlanguage}[2]{{%
\expandafter\ifx\csname l@#1\endcsname\relax
\typeout{** WARNING: IEEEtran.bst: No hyphenation pattern has been}%
\typeout{** loaded for the language `#1'. Using the pattern for}%
\typeout{** the default language instead.}%
\else
\language=\csname l@#1\endcsname
\fi
#2}}
\providecommand{\BIBdecl}{\relax}
\BIBdecl

\bibitem{lustig2008compressed}
{Lustig, Michael and Donoho, David L and Santos, Juan M and Pauly, John M},
  ``{Compressed sensing MRI},'' \emph{{IEEE signal processing magazine}},
  vol.~25, no.~2, pp. 72--82, 2008.

\bibitem{ye2019compressed}
{Ye, Jong Chul}, ``{Compressed sensing MRI: a review from signal processing
  perspective},'' \emph{{BMC Biomedical Engineering}}, vol.~1, no.~1, pp.
  1--17, 2019.

\bibitem{qu2012undersampled}
{Qu, Xiaobo and Guo, Di and Ning, Bende and Hou, Yingkun and Lin, Yulan and
  Cai, Shuhui and Chen, Zhong}, ``{Undersampled MRI reconstruction with
  patch-based directional wavelets},'' \emph{{Magnetic resonance imaging}},
  vol.~30, no.~7, pp. 964--977, 2012.

\bibitem{block2007undersampled}
{Block, Kai Tobias and Uecker, Martin and Frahm, Jens}, ``{Undersampled radial
  MRI with multiple coils. Iterative image reconstruction using a total
  variation constraint},'' \emph{{Magnetic Resonance in Medicine}}, vol.~57,
  no.~6, pp. 1086--1098, 2007.

\bibitem{li2022artificial}
{Li, Cheng and Li, Wen and Liu, Chenyang and Zheng, Hairong and Cai, Jing and
  Wang, Shanshan}, ``{Artificial intelligence in multiparametric magnetic
  resonance imaging: A review},'' \emph{{Medical Physics}}, 2022.

\bibitem{wang2021deep}
{Wang, Shanshan and Xiao, Taohui and Liu, Qiegen and Zheng, Hairong}, ``{Deep
  learning for fast MR imaging: a review for learning reconstruction from
  incomplete k-space data},'' \emph{{Biomedical Signal Processing and
  Control}}, vol.~68, p. 102579, 2021.

\bibitem{wang2021review}
{Wang, Shanshan and Cao, Guohua and Wang, Yan and Liao, Shu and Wang, Qian and
  Shi, Jun and Li, Cheng and Shen, Dinggang}, ``{Review and Prospect:
  Artificial Intelligence in Advanced Medical Imaging},'' \emph{{Frontiers in
  Radiology}}, vol.~1, p. 781868, 2021.

\bibitem{akccakaya2019scan}
{Ak{\c{c}}akaya, Mehmet and Moeller, Steen and Weing{\"a}rtner, Sebastian and
  U{\u{g}}urbil, K{\^a}mil}, ``{Scan-specific robust artificial-neural-networks
  for k-space interpolation (RAKI) reconstruction: Database-free deep learning
  for fast imaging},'' \emph{{Magnetic resonance in medicine}}, vol.~81, no.~1,
  pp. 439--453, 2019.

\bibitem{eo2018kiki}
{Eo, Taejoon and Jun, Yohan and Kim, Taeseong and Jang, Jinseong and Lee,
  Ho-Joon and Hwang, Dosik}, ``{KIKI-net: cross-domain convolutional neural
  networks for reconstructing undersampled magnetic resonance images},''
  \emph{{Magnetic resonance in medicine}}, vol.~80, no.~5, pp. 2188--2201,
  2018.

\bibitem{schlemper2017deep}
{Schlemper, Jo and Caballero, Jose and Hajnal, Joseph V and Price, Anthony and
  Rueckert, Daniel}, ``{A deep cascade of convolutional neural networks for MR
  image reconstruction},'' in \emph{{International conference on information
  processing in medical imaging}}.\hskip 1em plus 0.5em minus 0.4em\relax
  Springer, 2017, pp. 647--658.

\bibitem{zhang2018ista}
{Zhang, Jian and Ghanem, Bernard}, ``{ISTA-Net: Interpretable
  optimization-inspired deep network for image compressive sensing},'' in
  \emph{{Proceedings of the IEEE Conference on Computer Vision and Pattern
  Recognition (CVPR)}}, 2018, pp. 1828--1837.

\bibitem{sun2016deep}
{Sun, Jian and Li, Huibin and Xu, Zongben and others}, ``{Deep ADMM-Net for
  compressive sensing MRI},'' \emph{{Advances in neural information processing
  systems}}, vol.~29, 2016.

\bibitem{aggarwal2018modl}
{Aggarwal, Hemant K and Mani, Merry P and Jacob, Mathews}, ``{MoDL: Model-based
  deep learning architecture for inverse problems},'' \emph{{IEEE transactions
  on medical imaging}}, vol.~38, no.~2, pp. 394--405, 2018.

\bibitem{hammernik2018learning}
{Hammernik, Kerstin and Klatzer, Teresa and Kobler, Erich and Recht, Michael P
  and Sodickson, Daniel K and Pock, Thomas and Knoll, Florian}, ``{Learning a
  variational network for reconstruction of accelerated MRI data},''
  \emph{{Magnetic resonance in medicine}}, vol.~79, no.~6, pp. 3055--3071,
  2018.

\bibitem{yaman2020self}
{Yaman, Burhaneddin and Hosseini, Seyed Amir Hossein and Moeller, Steen and
  Ellermann, Jutta and U{\u{g}}urbil, K{\^a}mil and Ak{\c{c}}akaya, Mehmet},
  ``{Self-supervised learning of physics-guided reconstruction neural networks
  without fully sampled reference data},'' \emph{{Magnetic resonance in
  medicine}}, vol.~84, no.~6, pp. 3172--3191, 2020.

\bibitem{zbontar2018fastmri}
{Zbontar, Jure and Knoll, Florian and Sriram, Anuroop and Murrell, Tullie and
  Huang, Zhengnan and Muckley, Matthew J and Defazio, Aaron and Stern, Ruben
  and Johnson, Patricia and Bruno, Mary and others}, ``{fastMRI: An open
  dataset and benchmarks for accelerated MRI},'' \emph{{arXiv preprint
  arXiv:1811.08839}}, 2018.

\end{thebibliography}

\end{document}